\documentclass[aps,prl,twocolumn,superscriptaddress]{revtex4}
\usepackage{graphicx}
\usepackage{amsmath}
\usepackage{amssymb}
\usepackage{colordvi}
\usepackage{mathrsfs}
\usepackage{bm}
\usepackage{verbatim}
\usepackage{dcolumn}
\usepackage{bm}
\usepackage{epsfig}
\usepackage{subfigure}

\begin{document}
\title{Enhanced Robustness of Zero-line Modes in Graphene via a Magnetic Field}
\author{Ke Wang}
\affiliation{ICQD, Hefei National Laboratory for Physical Sciences at Microscale, and Synergetic Innovation Centre of Quantum Information and Quantum Physics, University of Science and Technology of China, Hefei, Anhui 230026, China}
\affiliation{CAS Key Laboratory of Strongly-Coupled Quantum Matter Physics and Department of Physics, University of Science and Technology of China, Hefei, Anhui 230026, China}
\author{Tao Hou}
\affiliation{ICQD, Hefei National Laboratory for Physical Sciences at Microscale, and Synergetic Innovation Centre of Quantum Information and Quantum Physics, University of Science and Technology of China, Hefei, Anhui 230026, China}
\affiliation{CAS Key Laboratory of Strongly-Coupled Quantum Matter Physics and Department of Physics, University of Science and Technology of China, Hefei, Anhui 230026, China}
\author{Yafei Ren}
\affiliation{ICQD, Hefei National Laboratory for Physical Sciences at Microscale, and Synergetic Innovation Centre of Quantum Information and Quantum Physics, University of Science and Technology of China, Hefei, Anhui 230026, China}
\affiliation{CAS Key Laboratory of Strongly-Coupled Quantum Matter Physics and Department of Physics, University of Science and Technology of China, Hefei, Anhui 230026, China}
\author{Zhenhua Qiao}
\email[Correspondence author:~]{qiao@ustc.edu.cn}
\affiliation{ICQD, Hefei National Laboratory for Physical Sciences at Microscale, and Synergetic Innovation Centre of Quantum Information and Quantum Physics, University of Science and Technology of China, Hefei, Anhui 230026, China}
\affiliation{CAS Key Laboratory of Strongly-Coupled Quantum Matter Physics and Department of Physics, University of Science and Technology of China, Hefei, Anhui 230026, China}
\begin{abstract}
  Motivated by recent experimental results for zero-line modes (ZLMs) in a bilayer graphene system [Nature Nanotechnol. \textbf{11}, 1060 (2016)], we systematically studied the influence of a magnetic field on ZLMs and demonstrated the physical origin of the enhanced robustness by employing nonequilibrium Green's functions and the Landauer--B\"{u}ttiker formula. We found that a perpendicular magnetic field can separate the wavefunctions of the counter-propagating kink states into opposite directions. Specifically, the separation vanishes at the charge neutrality point. The separation increases as the Fermi level deviates from the charge neutrality point and can reach a magnitude comparable to the wavefunction spread at a moderate field strength. Such spatial separation of oppositely propagating ZLMs effectively suppresses backscattering. Moreover, the presence of a magnetic field enlarges the bulk gap and suppresses the bound states, thereby further reducing the scattering. These mechanisms effectively increase the mean free paths of the ZLMs to approximately 1~$\mu$m in the presence of a disorder.
\end{abstract}
\maketitle

Recently, topologically nontrivial phases have attracted much attention because of their theoretical novelty and the potential applications in dissipationless electronics of their robust edge states, which are topologically protected from backscattering~\cite{edge1,edge2,edge3,edge4,edge5,edge6}. For the quantum Hall effect and the quantum anomalous Hall effect, the edge states are chiral with oppositely propagating edge states distributed at two boundaries of the sample. The spatial separation of counter-propagating states strongly suppresses backscattering against any type of impurity~\cite{QHE1,QHE2,QAHE1}. For the quantum spin-Hall effect, the edge states are spin-helical and only robust against elastic backscattering from nonmagnetic impurities because of the topological protection by the time-reversal invariance~\cite{QSHE}. Although these edge states are robust and have been realized in experiments, their practical application is difficult because of the requirement of strong magnetic field for quantum Hall effect or the introduction of both spin-orbit coupling and ferromagnetism for quantum anomalous Hall effect.

Different from the above symmetry-protected edge states, for the quantum valley Hall effect, although there is no rigorous bulk-edge correspondence, zero-line modes (ZLMs), also known as kink states, appear along the interface between regions with different valley topologies~\cite{edge3}. These helical ZLMs are protected by their large momentum separation. Because of the large wavefunction expansion, these modes are still quite robust against disorders and exhibit zero bend resistance~\cite{zlm}. More importantly, these one-dimensional ZLMs are highly tunable. By applying a spatially varying electric field to a chiral stacking bilayer graphene system, one can construct a single zero line, crossing double zero lines, and even zero line networks, exhibiting promising potential for applications in low-energy-consumption electronics~\cite{network, prl112}. Recently, a single zero and crossing double zero lines have been realized experimentally in bilayer graphene systems through careful gate alignment~\cite{nature nano, Natphys} and in samples with stacking line defects~\cite{Nature, nature nano}.

Inspired by the experimental results that suggest strong enhancement of the robustness of ZLMs by a magnetic field, herein, we investigate the transport properties of ZLMs in gated BLG systems under a perpendicular magnetic field to reveal the underlying physical origin. By varying the strength of the magnetic field, we systematically study the evolution of the electronic structure and the ZLM wavefunctions. We find that the presence of a magnetic field enlarges the bulk band gap and preserves the ZLM. However, the wavefunctions of counter-propagating ZLMs encoding with different valleys are pushed apart. The spatial separation increases with the magnetic-field strength. Moreover, at a given magnetic-field strength, spatial separation exhibits strong dependence on the Fermi energy. By decreasing the Fermi level from above the charge neutrality point, spatial separation decreases, vanishes at the charge neutrality point, and then increases again while in opposite direction as the Fermi level is decreased further. The spatial separation of counter-propagating wavefunctions strongly suppresses backscattering and makes the ZLMs more robust, as confirmed by our numerical calculations of conductances in the presence of a disorder. These results suggest that the presence of a magnetic field not only increases the robustness of ZLMs but also makes it possible to tune the transport by varying the Fermi level electrically.

\begin{figure}
	\centering
	\includegraphics[width=8.5cm,angle=0]{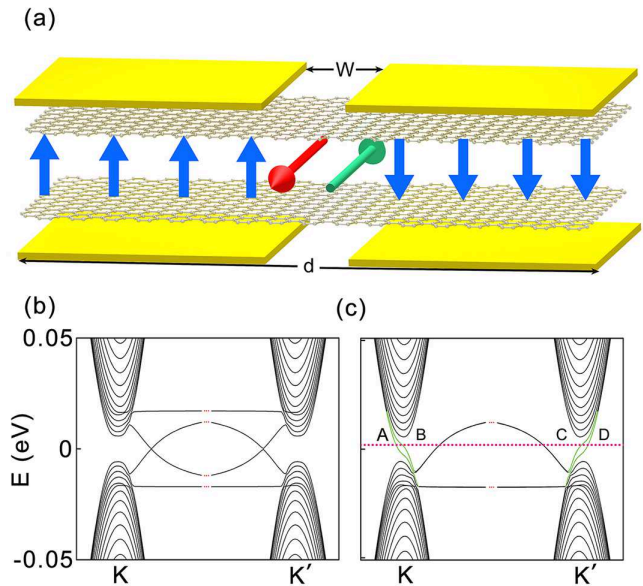}
	\caption{(a) Schematic of a dual-split-gated bilayer graphene (BLG) device. The blue arrows show the direction of the electric field. The green and red arrows correspond to modes that carry valley indexes K and K$^\prime$, respectively. (b) Electronic structure of the junction in the device shown in (a) with a zigzag boundary; the voltages in the adjoining regions are the same. Certainly, only the edge states appear. (c) The voltages in the adjoining regions are the opposite and kink states (marked in green) appear in the bulk gap.}\label{Figure1}
\end{figure}

The quantum valley Hall effect can be realized in AB-stacked bilayer graphene by applying a perpendicular electric field. When opposite electric fields are applied in neighboring regions, as illustrated in Fig.~\ref{Figure1}(a), where a downward (upward) electric field is applied in the region on the left (right), as denoted by the blue arrows. In the presence of an Anderson-type disorder, the $\pi$-orbital tight-binding Hamiltonian of this system can be expressed as
\begin{eqnarray}
H&=&-t \sum_{\langle ij \rangle} (c_i^{\dag} c_j+\text{h.c.}) -t_{\perp}\sum_{\langle{i\in{T},j\in {B}}\rangle } (c_i^{\dag} c_j+\text{h.c.}) \nonumber \\
&+& \sum_{i \in T} U_i c_i^{\dag} c_i - \sum_{i \in B} U_i c_i^{\dag} c_i +\sum_{ i } \epsilon_{i} c_i^{\dag} c_i,
\label{eq0}
\end{eqnarray}
where $c_i^{\dag}$ ($c_i$) is the electron creation (annihilation) operator on site~$i$. The first and second terms represent the intralayer and interlayer nearest-neighbor hopping, respectively, with hopping amplitudes of $t = 2.6$~eV and $t_{\perp} = 0.34$~eV. The third and fourth terms indicate the site energies at the top and bottom layers, respectively. At the left (right) part, the on-site energy $U_i$ is constant at $+U$ ($-U$). At the interface region of width $w$, we use a cosinusoidal potential profile to connect the two sides. The last term represents the on-site Anderson disorder, with $\epsilon_{i}$ being distributed randomly in the energy interval of $[-W/2, W/2]$, where $W$ measures the disorder strength. In our calculations, we set the system width to $d = 127.8$~nm and the interface region width to $w = 85.2$~nm. Note that we neglect the spin degree of freedom because the quantum valley Hall effect only involves the orbital degree of freedom of an electron.

In the presence of a perpendicular magnetic field $\mathbf{B}= \nabla \times \mathbf{A}$, the tight-binding Hamiltonian model is modified by introducing the Peierls phase in the hopping terms as follows:
\begin{equation}\label{eqution3}
t_{ij} \rightarrow t_{ij}{\rm e}^{-{\rm {i}}\frac{e}{\hbar} \int \mathbf{A}\cdot {\rm d} \boldsymbol{l}},
\end{equation}
where $\int \mathbf{{A}}\cdot {\rm d} \boldsymbol{l}$ is the integral of the vector potential from site~$j$ to site~$i$. In our calculations, we adopt the Landau gauge of $\mathbf{A} = -By\boldsymbol{e}_x$ for the perpendicular magnetic field $\mathbf{B} = B\boldsymbol{e}_z$.

With this $\pi$-orbital tight-binding Hamiltonian, we further study the ZLM transport properties using a two-terminal Landauer--B\"{u}ttiker formalism based on the Green's function technique~\cite{green}
\begin{eqnarray}
G_{\rm{RL}}=\frac{2e^2}{h}{\rm{Tr}}[\Gamma_{\rm R}G^{{r}}\Gamma_{\rm L}{G}^{{a}}],
\end{eqnarray}
where ${G}^{{r(a)}}$ is the retarded (advanced) Green's function of the central scattering region. $\Gamma_{\rm{R(L)}}= i[\Sigma^{{r}}_{{\rm R(L)}}-\Sigma^{{a}}_{{\rm R(L)}}]$ is the line width function, which describes the coupling between the right (left) lead and the central scattering region. $\Sigma^{r(a)}$ is the retarded (advanced) self-energy of the half-infinite leads calculated using the variant transfer matrix method~\cite{transfer}. In our calculations, we set the Fermi level to be inside the band gap. In the presence of a disorder, each data point is obtained by collecting over 30 samples with different disorder configurations.

In the absence of a disorder, the system exhibits translational symmetry along the zero line direction, allowing us to obtain the electronic structure of the system by directly diagonalizing the tight-binding Hamiltonian. When the applied electric fields in the left and right regions are opposite to each other (as shown in Fig.~\ref{Figure1}(a)), counter-propagating topological ZLMs appear at the interfacial region of width $w$ (as indicated by the red and green arrows). With a layer potential difference of $2U = 0.04$~eV, we calculate the band structure in the absence of a magnetic field. The results are shown in Fig.~\ref{Figure1}(c) wherein gapless ZLMs (marked in green) appear in valleys K and K$^\prime$. We note that the states in the two valleys exhibit opposite group velocities. Apart from the ZLMs, there are also some edge states (marked in gray) inside the band gap, which are localized at the sample boundaries. In contrast, when the electric fields in the left and right sides point in the same direction, the gapless ZLMs disappear, as shown in Fig.~\ref{Figure1}(b).

\begin{figure}
	\centering
	\includegraphics[width=8.5cm,angle=0]{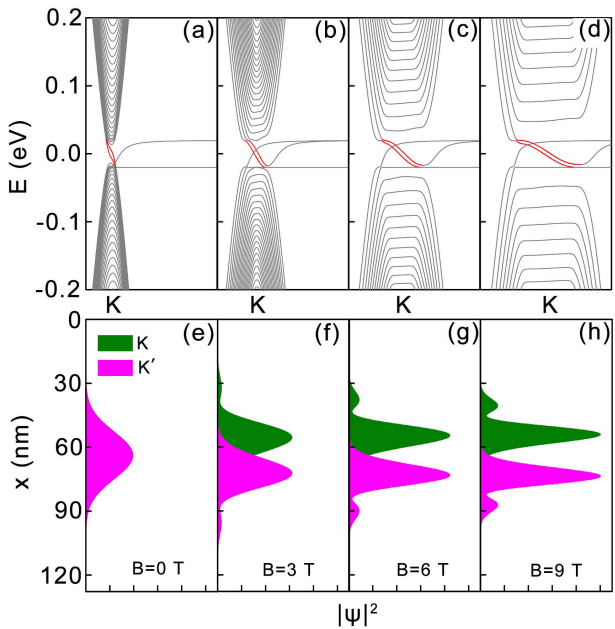}
	\caption{Band structure of the BLG line junction at different magnetic fields $B = 0$ (a), 3 (b), 6 (c), and 9~T (d). The device length is $d = 127.8$~nm, the junction width is $w = 85.2$~nm, and the layer potential difference is $2U = 40$~meV. The formation of the Landau levels lifts the bound states away from the energy range of the kink states (red). (e)--(h) Corresponding wavefunction distributions of the kink states in valleys K and K$^\prime$ (modes A and D, respectively, labeled in Fig.~\ref{Figure1} (c)), showing increasing separation as $B$ increases.}\label{Figure2}
\end{figure}

In the following discussion, we focus on the case of opposite electric fields in the two sides of the interface. In this case, there are four ZLMs inside the band gap (as denoted by A, B, C, and D in Fig.~\ref{Figure1}(c)) with Fermi energy $E_{\rm F} = 4$~meV. Figure~\ref{Figure2}(a) shows a magnified view of the band structure in valley K. The wavefunctions of A and D, which are located in valleys K and K$^\prime$, respectively, are shown in Fig.~\ref{Figure2}(e), where we find that the wavefunctions from both valleys coincide with each other. In the presence of a magnetic field, the band structure is modified as shown in Figs.~\ref{Figure2}(b)--(d). From these figures, we find that the presence of a magnetic field enhances the bulk band gap and lifts the bulk bands away from the ZLMs plotted in red. As the band gap increases, the ZLM wavefunctions become narrowed, as shown in Figs.~\ref{Figure2}(f)--(h). Moreover, we find that the presence of a magnetic field drives the ZLM wavefunction in valley K upward and that in valley K$^\prime$ downward, thereby reducing the overlap between counter-propagating states.

\begin{figure}
	\centering
	\includegraphics[width=8.5cm, angle=0]{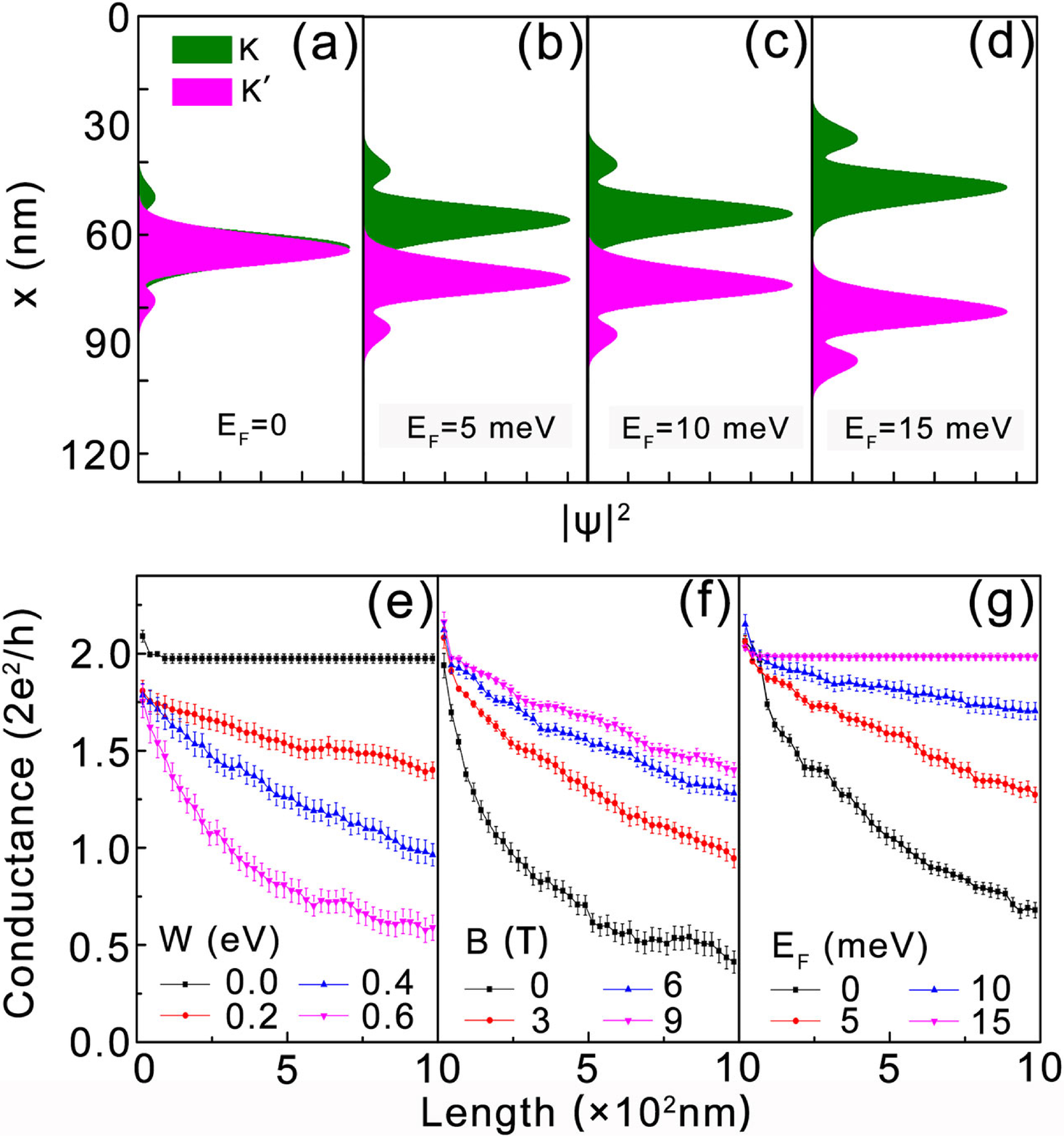}
	\caption{(a)--(d) Corresponding wavefunction distributions of the kink states in valleys K and K$^\prime$ (modes A and D, respectively, labeled in Fig.~\ref{Figure1}(c)) at different Fermi energies $E_F = 0$ (a), 5 (b), 10 (c), and 15~meV (d) but the same magnetic field strength $B = 9$~T.
	(e) and (f) Average conductance vs. device length under different parameter strengths.
	In (e), the Fermi level is $E_{\rm F} = 10$~meV, there is no magnetic field, and the disorder strength is varied as $W = 0$, 200, 400, and 600~meV. The band gap is set as $2U = 40$~meV.
	In (f), the Fermi level is $E_{\rm F} = 6$~meV, the magnetic field strength is varied as $B = 0$, 3, 6, and 9~T, and the disorder strength is $W = 600$~meV.
	In (g), the magnetic field strength is $B = 9$~T, the Fermi level is varied as $E_{\rm F} = 0$, 5, 10, and 15~meV, and the disorder strength is $W = 60$~meV.
	Each point is an average of 30 samples with different disorder configurations.}\label{Figure3}
\end{figure}

Furthermore, in the presence of a magnetic field, the wavefunction distribution can be tuned by varying the Fermi energy electrically, as shown in Figs.~\ref{Figure3}(a)--(d). In these figures, the magnetic field is set to be 3~T. By increasing the Fermi energy from $E_{\rm F} = 0$~meV to $E_{\rm F} = 15~meV$, we find that the ZLM wavefunction in valley K (K$^\prime$) moves upward (downward). At the charge neutrality point of $E_f = 0$~meV, the wavefunctions almost coincide with each other.

In brief, the presence of a magnetic field not only separates the ZLMs and bulk states by a larger band gap but also spatially separates the counter-propagating ZLMs in different valleys. Both effects can effectively suppress the backscattering of the ZLMs. Thus, the presence of a magnetic field is expected to strongly enhance the robustness of the ZLMs. To demonstrate the influence of the magnetic field, Fermi energy, and disorder, we also study the ZLM conductance for junctions of different lengths. We begin by studying the case that is in the absence of a magnetic field, as shown in Fig.~\ref{Figure3}(e), where we find a quantized conductance that is independent of the junction length in the absence of a disorder. As the disorder strength is increased, the conductance decreases with an increase in the system length. For a disorder strength of $W = 600$~meV, the conductance decreases to a value of $0.5~e^2/h$ for a junction length of 1~$\mu$m. At the same disorder strength, the conductance is increased by applying a magnetic field, as shown in Fig.~\ref{Figure3}(f). In the presence of a magnetic field of 9~T, the conductance of the 1-$\mu$m junction is increased by a factor of three. The enhancement of the conductance by the magnetic field becomes increasingly obvious as the Fermi energy deviates from the charge neutrality point, as shown in Fig.~\ref{Figure3}(g), under a magnetic field of 9~T. In this figure, we find that at the charge neutrality point of $E_{\rm F} = 0$~meV, even under a magnetic field of 9~T, the conductance decreases to approximately $0.5~$e$^2$/h. As the Fermi energy is gradually increased, the influence of the disorder on the conductivity becomes weaker and a quantized conductance of a sample as long as 1~$\mu$m can be realized when the Fermi energy reaches $0.015$~eV.

\begin{figure}
	\centering
	\includegraphics[width=8.5cm,angle=0]{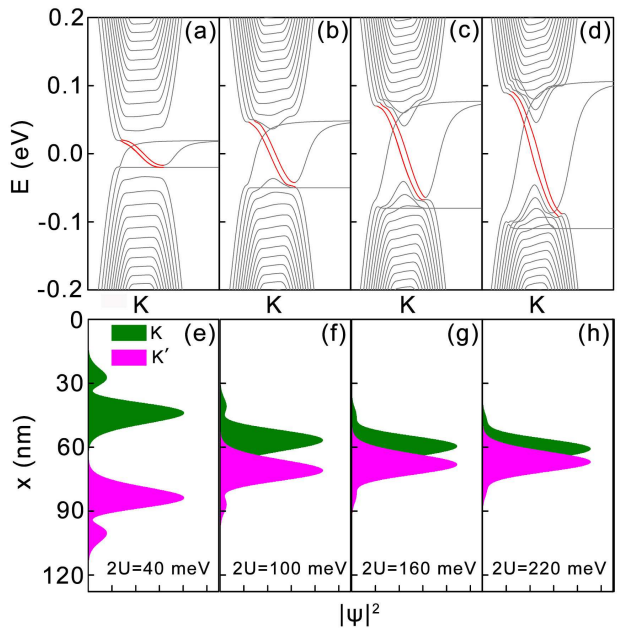}
	\caption{(a)--(d) Band structures of the BLG line junction calculated with $2U = 40$ (a), 100 (b), 160 (c), and 220~meV (d), a magnetic field strength of $B = 6$~T, and a Fermi level of $E_{\rm F} = 12$~meV. Increasing $U$ increases the energy range in which only the kink states (marked in red) exist. (e)--(h) Corresponding wavefunction distributions of the kink states in valleys K and K$^\prime$ (modes A and D, respectively, labeled in Fig.~\ref{Figure1}(c)), showing increasing separation as the layer potential difference $U$ is decreased.}\label{Figure4}
\end{figure}

Nevertheless, the influence of the magnetic field reduces as the band gap increases. In Figs.~\ref{Figure4}(a)--(d), we plot the band structure for differing layer potential difference $U$ with the same magnetic field strength and Fermi level. We find that the band gap increases with the layer potential difference. The ZLM wavefunctions are also shown in Figs.~\ref{Figure4}(e)--(h). We find that the spatial separation becomes considerably increasing with decrease of the band gap. This is because increasing the layer potential difference enhances the electron confinement, and the wavefunctions are restricted to the sharp interface between the left and right regions, meaning that there is no room for the wavefunctions to move.

In summary, we systematically studied the effect of a magnetic field on the ZLM transport properties at different band gaps and Fermi energies. We found that the presence of a magnetic field not only enlarges the band gap, separating the ZLMs from the bulk states by a larger energy gap, but also spatially separates the wavefunction distributions of counter-propagating ZLMs. Furthermore, we found that decreasing the layer potential difference increases wavefunction separation. Both effects strongly enhance the robustness of ZLMs against a disorder. By calculating the ballistic length of the ZLMs under different disorder strengths, we numerically confirmed that the presence of a magnetic field can effectively enlarge the ballistic length. Moreover, we found that the spatial separation of the counter-propagating ZLMs depends strongly on the Fermi energy. Under a moderate magnetic field, increasing the Fermi energy from the CNP of $E_{\rm F} = 0$~meV increases the distance between counter-propagating ZLMs. Such an effect makes the ZLM transport property highly tunable by a magnetic field. These findings provide a new strategy for enhancing the robustness of ZLMs.

This work was financially supported by the National Key Research and Development Program (Grant Nos.\ 2016YFA0301700, 2017YFB0405703), the China Government Youth 1000-Plan Talent Program, and the NNSFC (Grant No.\ 11474265). We are grateful to the Supercomputing Center of USTC for providing high-performance computing assistance.

\end{document}